\documentclass[runningheads]{llncs}
\usepackage[T1]{fontenc}
\usepackage{textcomp}
\usepackage{graphicx}
\usepackage{hyperref}
\usepackage[strings]{underscore}
\usepackage{enumitem}
\usepackage{booktabs}
\usepackage[caption=false,font=normalsize,labelfont=sf,textfont=sf]{subfig}
\usepackage{listings, xcolor, chngcntr}
\usepackage{tabularx}
\colorlet{punct}{red!60!black}
\definecolor{background}{rgb}{.97,.97,.97}
\definecolor{delim}{RGB}{20,105,176}
\colorlet{numb}{magenta!60!black}

\lstdefinelanguage{json}{
    backgroundcolor=\color{background},
    literate=
     *{0}{{{\color{numb}0}}}{1}
      {1}{{{\color{numb}1}}}{1}
      {2}{{{\color{numb}2}}}{1}
      {3}{{{\color{numb}3}}}{1}
      {4}{{{\color{numb}4}}}{1}
      {5}{{{\color{numb}5}}}{1}
      {6}{{{\color{numb}6}}}{1}
      {7}{{{\color{numb}7}}}{1}
      {8}{{{\color{numb}8}}}{1}
      {9}{{{\color{numb}9}}}{1}
      {:}{{{\color{punct}{:}}}}{1}
      {,}{{{\color{punct}{,}}}}{1}
      {\{}{{{\color{delim}{\{}}}}{1}
      {\}}{{{\color{delim}{\}}}}}{1}
      {[}{{{\color{delim}{[}}}}{1}
      {]}{{{\color{delim}{]}}}}{1},
}

\definecolor{verylightgray}{rgb}{.97,.97,.97}

\lstdefinelanguage{Solidity}{
	keywords=[1]{anonymous, assembly, assert, balance, break, call, callcode, case, catch, class, constant, continue, constructor, contract, debugger, default, delegatecall, delete, do, else, emit, event, experimental, export, external, false, finally, for, function, gas, if, implements, import, in, indexed, instanceof, interface, internal, is, length, library, log0, log1, log2, log3, log4, memory, modifier, new, payable, pragma, private, protected, public, pure, push, require, return, returns, revert, selfdestruct, send, solidity, storage, struct, suicide, super, switch, then, this, throw, true, try, typeof, using, value, view, while, with, addmod, ecrecover, keccak256, mulmod, ripemd160, sha256, sha3}, 
	keywordstyle=[1]\color{blue}\bfseries,
	keywords=[2]{address, bool, byte, bytes, bytes1, bytes2, bytes3, bytes4, bytes5, bytes6, bytes7, bytes8, bytes9, bytes10, bytes11, bytes12, bytes13, bytes14, bytes15, bytes16, bytes17, bytes18, bytes19, bytes20, bytes21, bytes22, bytes23, bytes24, bytes25, bytes26, bytes27, bytes28, bytes29, bytes30, bytes31, bytes32, enum, int, int8, int16, int24, int32, int40, int48, int56, int64, int72, int80, int88, int96, int104, int112, int120, int128, int136, int144, int152, int160, int168, int176, int184, int192, int200, int208, int216, int224, int232, int240, int248, int256, mapping, string, uint, uint8, uint16, uint24, uint32, uint40, uint48, uint56, uint64, uint72, uint80, uint88, uint96, uint104, uint112, uint120, uint128, uint136, uint144, uint152, uint160, uint168, uint176, uint184, uint192, uint200, uint208, uint216, uint224, uint232, uint240, uint248, uint256, var, void, ether, finney, szabo, wei, days, hours, minutes, seconds, weeks, years},	
	keywordstyle=[2]\color{teal}\bfseries,
	keywords=[3]{block, blockhash, coinbase, difficulty, gaslimit, number, timestamp, msg, data, gas, sender, sig, value, now, tx, gasprice, origin},	
	keywordstyle=[3]\color{violet}\bfseries,
	identifierstyle=\color{black},
	sensitive=false,
	comment=[l]{//},
	morecomment=[s]{/*}{*/},
	commentstyle=\color{gray}\ttfamily,
	stringstyle=\color{red}\ttfamily,
	morestring=[b]',
	morestring=[b]"
}

\makeatletter
\newcommand{\printfnsymbol}[1]{%
  \textsuperscript{\@fnsymbol{#1}}%
}
\makeatother
\begin{document}
\counterwithin{lstlisting}{section}
\title{Distributed Ledger for Provenance Tracking of Artificial Intelligence Assets}
%
%
\author{Philipp L\"uthi\inst{1}\thanks{equal contribution} \and
Thibault Gagnaux\inst{1}\printfnsymbol{1} \and
Marcel Gygli\inst{1}}
\authorrunning{Philipp L\"uthi, Thibault Gagnaux, Marcel Gygli}
%
%
\institute{Fachhochschule Nordwestschweiz FHNW, Institut f\"ur Interaktive Technologien (IIT), Windisch, Switzerland\\
\email{\{firstname\}.\{lastname\}@fhnw.ch}\\
}
\maketitle              
\begin{abstract}
High availability of data is responsible for the current trends in Artificial Intelligence (AI) and Machine Learning (ML).
However, high-grade datasets are reluctantly shared between actors because of lacking trust and fear of losing control.
Provenance tracing systems are a possible measure to build trust by improving transparency.
Especially the tracing of AI assets along complete AI value chains bears various challenges such as trust, privacy, confidentiality, traceability, and fair remuneration.
In this paper we design a graph-based provenance model for AI assets and their relations within an AI value chain.
Moreover, we propose a protocol to exchange AI assets securely to selected parties.
The provenance model and exchange protocol are then combined and implemented as a smart contract on a permission-less blockchain.
We show how the smart contract enables the tracing of AI assets in an existing industry use case while solving all challenges.
Consequently, our smart contract helps to increase traceability and transparency, encourages trust between actors and thus fosters collaboration between them.
\end{abstract}
%
%
%

\section{Introduction}
\label{sec:intro}
Artificial intelligence (AI) is continuously becoming more critical for businesses.
As reported by the \textit{Big Data and AI Executive Survey 2019} 92\% of the firms are increasing their pace of investing in Artificial Intelligence and Big Data~\cite{thomash.davenport2019}. 
The availability of data drives AI development. 
Google, for example, surpassed 3 billion searches worldwide per day in 2012~\cite{googlellc2012}.
Moreover, data has replaced oil as the most valuable resource and is becoming the new currency of the digital era~\cite{economist2017}.

\begin{figure}[t]
  \centering
  \includegraphics[width=\linewidth]{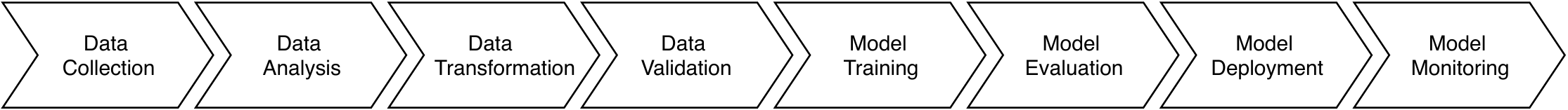}
  \caption{The eight phases in typical machine learning workflows. Each phase creates or adds value to assets that are transferred between different phases.}
  \label{fig:mlworkflow}
\end{figure}

Creating value from data consists of multiple steps. 
For Machine Learning (ML), Baylor et al. identified eight steps shown in Figure \ref{fig:mlworkflow}.
Between each of these phases, value is passed in the form of assets.
Value chains are created by linking these assets together, named AI value chain if all involved assets are relevant for the creation of an AI solution.
Machine or deep-learning that are used to solve a defined task such as medical image analysis~\cite{litjens2017} are examples of AI value chains. 

\begin{figure}[ht]
  \centering
  \includegraphics[width=\linewidth]{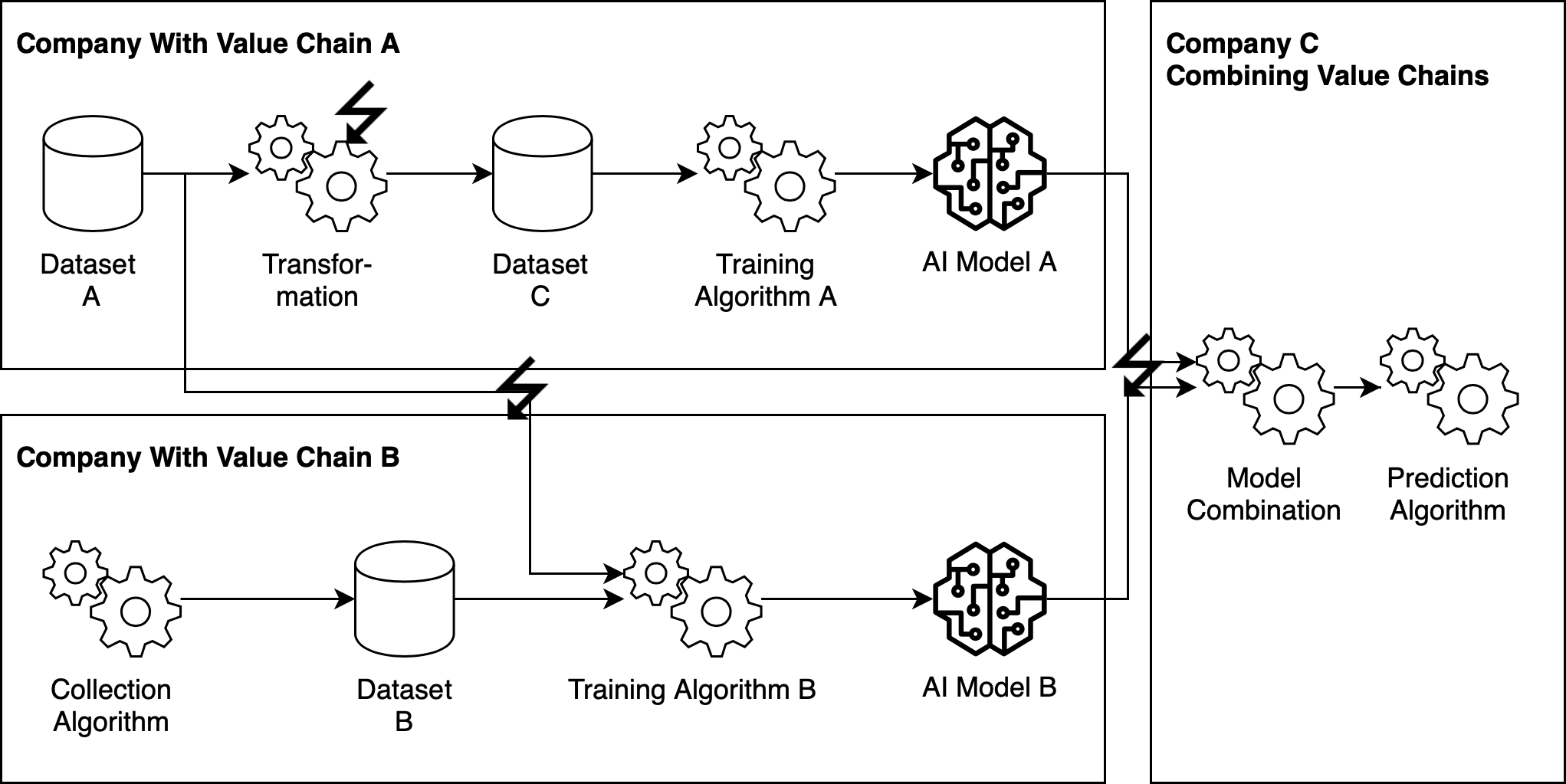}
  \caption{Coexisting and interacting AI value chains involving different companies. 
            Datasets and Models are shared between the three companies.
            Friction points where challenges arise are marked with the lightning symbol.
            Trust issues occur when handing over an AI asset from one company to another.
            Privacy and confidentiality concerns arise whenever data is processed within a value chain.
            Traceability and fair remuneration need to be addressed, as otherwise, companies will often not collaborate.}
  \label{fig:valuechains}
\end{figure}

Today, an AI value chain often involves experts from different organizations.
It is also possible that multiple value chains coexist or interact with each other as shown in Figure \ref{fig:valuechains}.
As a result, AI assets need to be exchanged between actors possibly merging or splitting value chains. 
The figure shows how three different companies exchange AI assets to create value.
Additionally, points of friction are shown where challenges arise.
These are the challenges we address within this work.

Participants of such AI value chains face several challenges when they try to exchange AI assets:
\begin{itemize}[leftmargin=0pt]
    \item[] \textbf{Trust} is one of the core issues that need to be addressed.
                Currently, when handing over an AI asset, the provider needs to trust the receiver as well as the medium of transfer because they lose sovereignty and control over the asset.
                This setting discourages the sharing of AI assets because there are no good systems that allow creating such trust between two parties that do not know each other.
    \item[] \textbf{Privacy}: From a provider's perspective, it is vital to know how their asset is used to prevent abuse.
            Especially if personal information is in the data, data regulations such as GDPR\footnote{See \url{https://eugdpr.org}} might apply.
            We do not aim at providing a solution towards data privacy within this work, but are interested in providing solutions such that data providers can keep track where data is being used.
            In regards to GDPR, this would potentially enable the retraction of data items from multiple AI value chains at once.
    \item[] \textbf{Confidentiality}: A provider has a key interest in keeping their assets either private or only visible to a selected number of other actors.
                This allows providers to protect their business interests and for example share sensitive data only with certified companies.
    \item[] \textbf{Traceability, Transparency \& Auditability}: If AI assets are exchanged without their provenance, the receiver may not have any possibility to verify their correctness. 
                By verifying the asset's correctness, errors that otherwise might propagate through the value chain unnoticed can be detected early.
    \item[] \textbf{Fair Remuneration}: A provider wants to know who is making use of their assets to claim their reward.
                From a receiver's perspective, it might be of interest to identify the involved actors to fairly compensate them.
\end{itemize}

Distributed Ledger Technology (DLT)~\cite{maull2017} allows creating a distributed database, spread across many nodes.
Blockchain~\cite{crosby2016a} is one of the most prominent concepts for the actual implementation of DLT.
In a blockchain all blocks are cryptographically linked with one another, depending on the data that is stored within them.
This ensures that data cannot be modified once it has been stored.

In this paper, we introduce a smart contract~\cite{clack2016} concept that allows tracking the provenance of AI assets along their value chains.
The smart contract can be deployed on the Ethereum blockchain~\cite{wood2014a}.
Data providers will be able to register their data using the smart contract and data consumers can register how they used this data and what kind of operations they applied on it.
This smart contract creates trust for all involved parties, as it allows for independent auditing of all registered transactions to verify its integrity.
We will show how this approach solves the privacy, auditability, and fair remuneration challenges.

The rest of the paper is structured as follows: Section \ref{sec:related} describes the approaches and solutions of other provenance systems and identifies a gap that we will address.
Section \ref{sec:provenancemodel} illustrates how we define our provenance model and how it solves the first three challenges privacy, auditability and fair remuneration. 
Section \ref{sec:implementaton} shows the functionality of the smart contract and the protocol specification that allow us to solve the remaining trust and confidentiality challenges.
We validate our solution using a real-world medical use case in Section \ref{sec:validation}.
The paper closes with a discussion and pointers towards future work.

\section{Related Work}
\label{sec:related}
In the following, we introduce existing approaches for the purpose of tracing the provenance of data in AI, and then existing provenance tracing solutions that are using blockchain technology.
We discuss these approaches, introduce the limitations of existing approaches and identify a gap that we address within this work.

\subsection{Provenance Tracing for Data in AI Development}
With the rise of Big Data, traditional data processing and provenance tracing~\cite{woodruff1997,buneman2001} became inapplicable. 
However, Provenance tracking has been identified as a key requirement for Big Data applications~\cite{labrinidis2012}. 
MapReduce~\cite{dean2008} is a specialized framework enabling parallel processing of high volume data.
Provenance capturing for MapReduce workflows is possible. 
Park et al. developed RAMP~\cite{park2011}, a provenance capturing system that extends \textit{Hadoop}\footnote{See \url{https://hadoop.apache.org}}.

In the area of machine learning, tracking the provenance of data points and training algorithms can be automated.
The importance of single data points can be calculated. 
Ma et al. designed LAMP~\cite{ma2017} to automate the partial derivative calculation of each data point evaluating its importance on the machine learning algorithm's result. 
Schelter et al. designed a system~\cite{schelter2017} that automatically extracts and stores provenance information of common artifacts in machine learning experiments. 
The system can be integrated into many popular machine learning frameworks to improve the reproducibility and comparability of machine learning experiments.

\subsection{Provenance Tracing Using Blockchain}
Blockchain technology is well suited in environments where trust between actors is needed as it makes a middleman obsolete. 
Additionally, storing provenance information on a blockchain is beneficial due to its immutable nature. 
Liang et al.~\cite{liang2017} introduce ProvChain, a cloud architecture that gathers and validates provenance data by inserting them into blockchain transactions.
Also, ProvChain provides security features such as tamper-proof provenance and user privacy.

Ramachandran and Kantarcioglu~\cite{ramachandran2018} use Ethereum to develop a secure and immutable scientific data provenance management framework called SmartProvenance that validates the provenance data using a use case-tailored verification script. 
During the verification, involved actors approve or reject proposed changes in a voting process.

ProvChain and SmartProvenance focus on tracking provenance on a single file and solve auditability by storing each file change on the blockchain.
As they have only one value chain they do not need to address trust or confidentiality.

Sarpatwar et al.~\cite{sarpatwar2019a} combined blockchain and AI and propose a concept for trusted AI. 
They illustrate the needed requirements and key blockchain constructs for trusted AI and demonstrate how these can be used to represent provenance using a federated learning~\cite{konecny2016} use case.
For their specific use case, they did not need to address the trust and confidentiality challenges.
They also do not address the issue of auditability of data.

\subsection{Limitations of State of the Art}
Existing solutions allow storing provenance of digital assets on central or distributed databases~\cite{woodruff1997,buneman2001}.
These databases are controlled by a single authority, making them vulnerable to untruthful modification.
This requires the trust of all involved parties into this central authority, which hinders collaboration. 

For Big Data and machine learning workflows, provenance tracking of individual data points with traditional databases is not feasible.
Therefore, specialized frameworks have been developed that allow tracing the provenance using MapReduce.

Blockchain technology allows storing provenance without the need for a centralized authority.
Consequently, all actors can participate equally and validate the transactions of other actors.
Furthermore, provenance information stored on the blockchain is immutable and therefore false modifications are impossible. 

Existing blockchain solutions can track the changes applied to single digital assets and AI models trained in a federated learning scenario.
They, however, cannot trace all phases of a typical AI development workflow. 
In particular, the exchange and transformation of AI assets between interacting AI value chains are not supported, as can be seen in Figure \ref{fig:valuechains}. For example, it is not possible to track how data is collected or transformed before it is used for AI model training. Furthermore, current blockchain solutions do not address how assets can be shared confidentially and selectively. The models trained in federated learning are directly stored on the blockchain and thus publicly accessible.

None of the abovementioned works cover all outlined challenges in a sufficient manner.
This calls for further research on how to design a system that addresses all of them.
In this work, we generalize the provenance model of Sarpatwar et al.~\cite{sarpatwar2019a} to be able to represent interacting AI value chains of any sort.
Additionally, we build a system that stores this provenance model and supports the exchange of confidential assets without the need for a centralized authority.

\section{Provenance Model for AI Assets on a Public Permission-less Blockchain}
\label{sec:provenancemodel}
Our provenance model extends the model of Sarpatwar et al.~\cite{sarpatwar2019a}, which differentiates between datasets, operations, and models.
This existing concept solves the privacy, auditability, and fair remuneration within the federated learning~\cite{konecny2016} context.
Thus, it has several limitations:
Datasets and models can exist only with a corresponding operation and the types of operations are finite. 
Operations always result in a model.
Trust is generated by making generated models as well as their coefficients public.
Confidentiality is addressed only for private datasets. 
Besides, their provenance model does not allow to track transformations on datasets and thus does not solve any challenges for general interacting value chains where an exchange of datasets is needed as illustrated in Figure \ref{fig:valuechains}. 

\begin{figure}[!t]
    \centering
    \subfloat[]{\includegraphics[width=.15\linewidth]{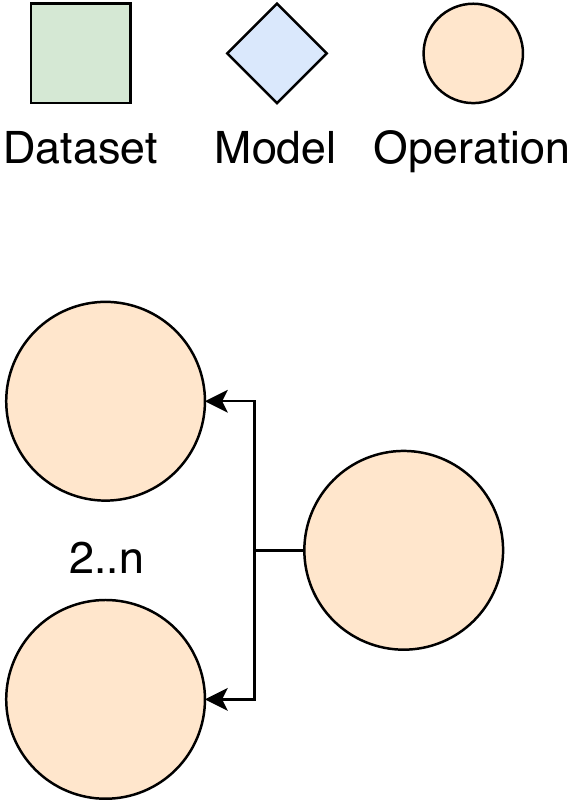}
    \label{fig:opcombination}}
    \hfil
    \subfloat[]{\includegraphics[width=.15\linewidth]{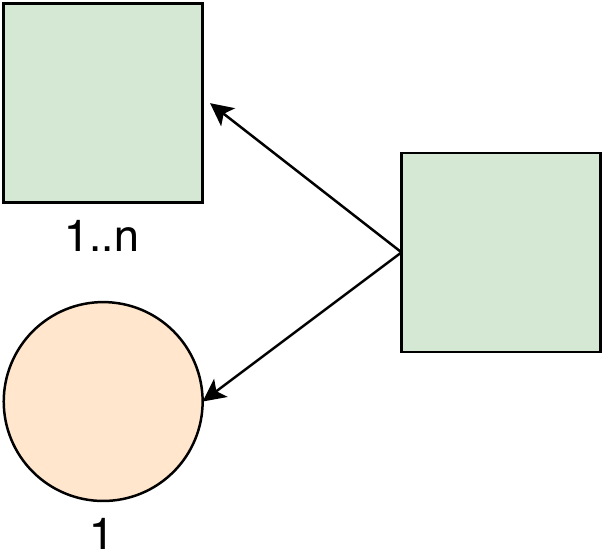}
    \label{fig:dstransformation}}
    \hfil
    \subfloat[]{\includegraphics[width=.15\linewidth]{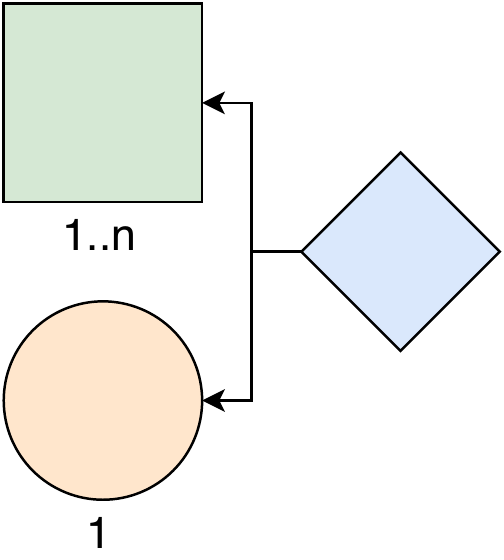}
    \label{fig:dsreduction}}
    \hfil
    \subfloat[]{\includegraphics[width=.15\linewidth]{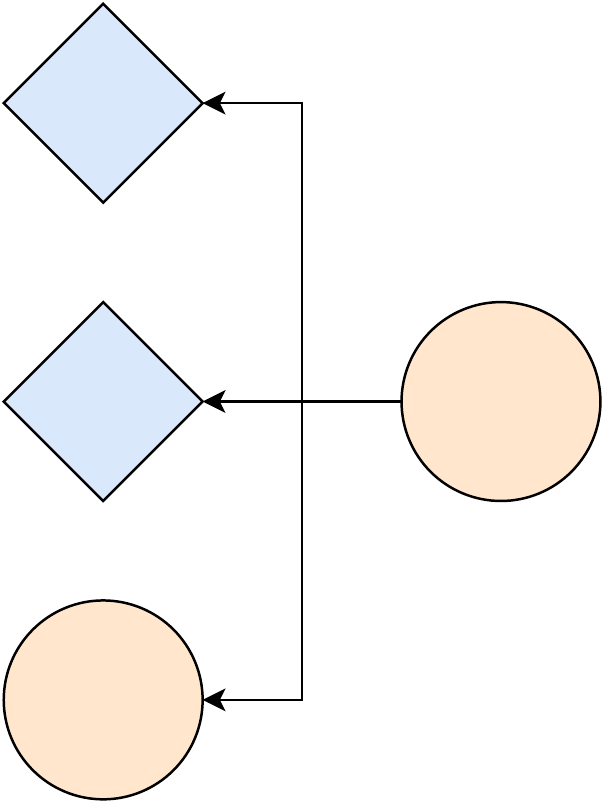}
    \label{fig:motraining}}
    \hfil
    \subfloat[]{\includegraphics[width=.15\linewidth]{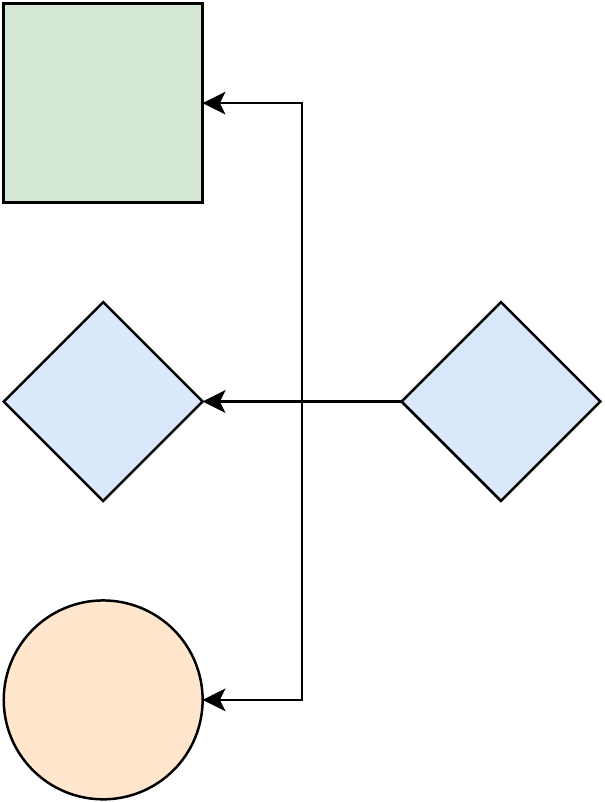}
    \label{fig:motransferlearning}}
    \hfil
    \subfloat[]{\includegraphics[width=.15\linewidth]{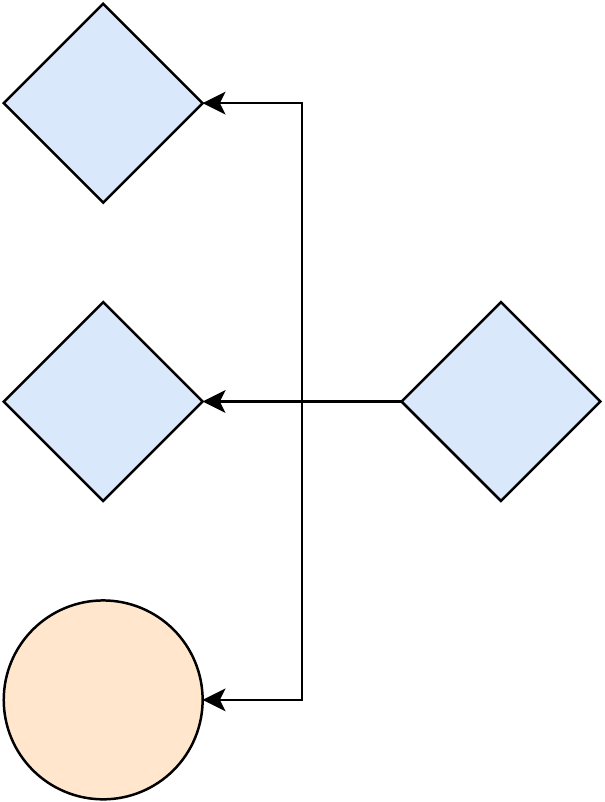}
    \label{fig:mofederatedlearning}}
    \hfil
    \caption{
      The basic building blocks can be used to combine Operations (a), transform Datasets (b), or create Models (c).
      By combining these types we are able to model concrete examples such as combining models to bring them into application (d), transfer learning (e), or the fusion part of federated learning (f).
      }
\end{figure}

Our goal is to generalize this provenance model and support interacting AI value chains and add the ability to track datasets and models without the need for the corresponding operation.
Furthermore, the provenance model will allow participants to define their operations. 
To achieve this goal we redefine the three types of AI assets and their relations as follows:
\begin{itemize}[leftmargin=0pt]
 \item[] \textbf{Operation}: An operation may represent any executable algorithm. 
                             In an AI value chain operations might be used for data collection, transformation, combination, reduction, analysis, training, etc. 
                             Multiple operations can be combined into one single operation as shown in Figure \ref{fig:opcombination}.
 \item[] \textbf{Dataset}: A dataset represents any composition of digital data. 
                           A dataset might be transformed (e.g. an anonymization algorithm) resulting in a new dataset as shown in Figure \ref{fig:dstransformation} or multiple datasets might be reduced (e.g. a filtering algorithm) to one dataset as shown in Figure \ref{fig:dsreduction}.
 \item[] \textbf{Model}: Combining an operation with a dataset (e.g. a classification algorithm) can result in a model as shown in Figure \ref{fig:motraining}. 
                        Additionally, combining a model, a dataset and an operation might result in a new model (e.g. transfer learning) as shown in Figure \ref{fig:motransferlearning}. 
                        Lastly, multiple models can be combined into one model (e.g., through federated learning) as shown in figure \ref{fig:mofederatedlearning}.
\end{itemize}

\begin{figure}[t]
    \centering
    \includegraphics[width=0.7\linewidth]{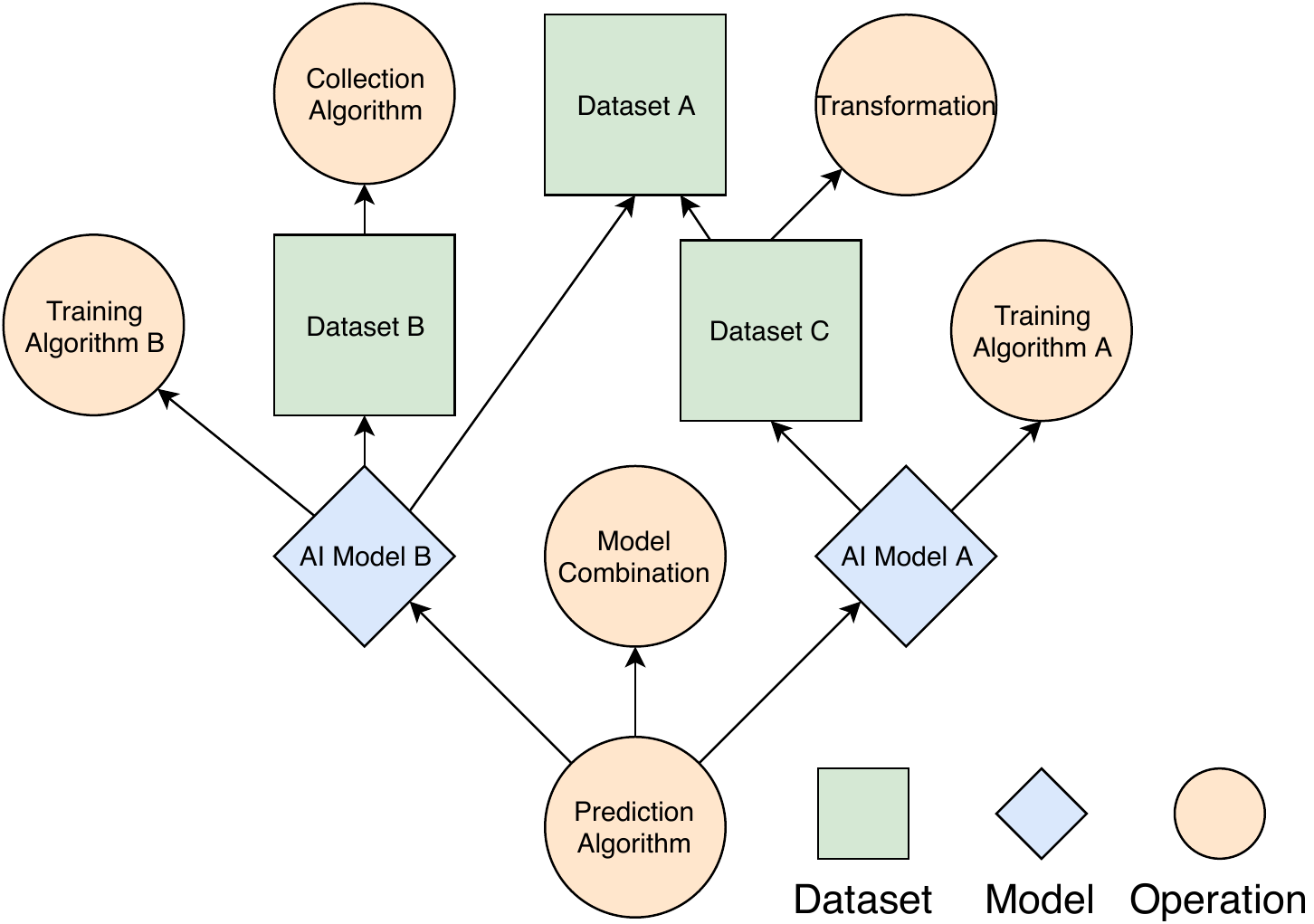}
    \caption{The provenance model created from the interacting value chains shown in Figure \ref{fig:valuechains}.
    The ``Collection Algorithm'' has a parent of relation with ``Dataset B'', and similarly the ``AI Model B'' has a child of relation with ``Dataset B''.}
    \label{fig:simpleprovenance}
\end{figure}

We represent the provenance model as a directed acyclic graph (DAG)~\cite{bondy1976} with nodes representing the AI assets.
Edges in this graph either represent a \textbf{Parent Of} or \textbf{Child Of} relationship between two assets.
Figure \ref{fig:simpleprovenance} shows how the interacting value chains of Figure \ref{fig:simpleprovenance} are transformed into this graph.
The ``Collection Algorithm'' is a parent of ``Dataset B'', and similarly ``AI Model B'' is a child of ``Dataset B''.

Using this graph representation, the traceability, fair remuneration, and privacy challenges can be solved using graph traversal algorithms.
This will be shown in detail in Section \ref{sub:protocol}.

\section{Implementation} 
\label{sec:implementaton}
In this section, we propose the implementation of the provenance model outlined in Section \ref{sec:provenancemodel} on a public blockchain. 
Our implementation introduces the possibility to exchange confidential AI assets between different actors, therefore addressing the confidentiality challenges. 
We use the Ethereum blockchain~\cite{wood2014a} as a deployment and execution framework.
In this framework, code is executed through so-called \textit{smart contracts}~\cite{buterin2014a}.
These smart contracts can write data in two different ways \textit{State Storage} or \textit{Logs}.
\textit{State Storage} stores information directly in the state of the smart contract and can be modified by it.
\textit{Logs} are a cheaper form of data storage that can not be read or modified by the smart contract.
When Logs are written to the blockchain they emit \textit{events} on the smart contract, for which client applications are able to listen to.

Additionally, we define a protocol to interact with our implementation.
Based on this protocol, we will show how we provide solutions for all challenges introduced in Section \ref{sec:intro}.
Every client that implements the protocol accordingly, helps to build and enforce provenance, therefore increasing trust into the registered AI assets.

\subsection{System Overview}
\label{sysOverview}
\begin{figure}[ht]
  \centering
  \includegraphics[width=0.75\linewidth]{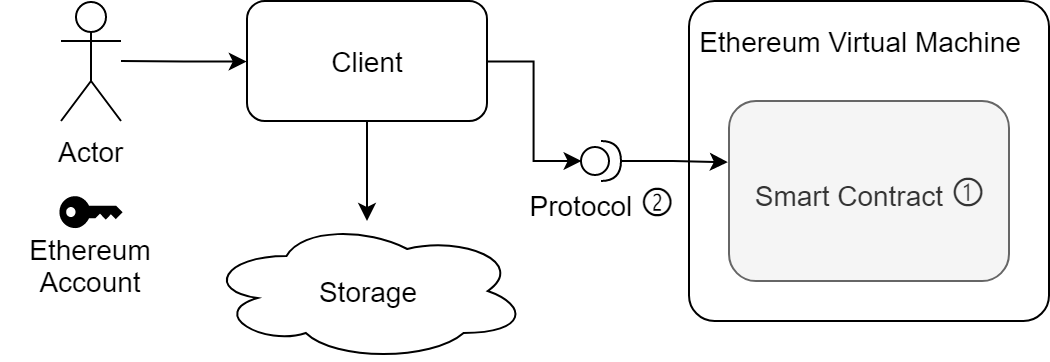}
  \caption{System context diagram: The complete system context with the described smart contract {\normalsize \textcircled{\small 1}} (see Section \ref{sub:smartcontract}) on the Ethereum Network and the specified protocol {\normalsize \textcircled{\small 2}} (see Section \ref{sub:protocol}).}
  \label{fig:systemOverview}
\end{figure}

Our system is comprised of a smart contract and a protocol specification to interact with it as shown in Figure \ref{fig:systemOverview}.
The smart contract writes the results of all actions performed on it into the blockchain.
By specifying what information is stored the contract enforces the provenance model from Section \ref{sec:provenancemodel}.
The protocol specification defines how to work with the smart contract in such a way that the stored information can be retrieved and the provenance model can be built.
Furthermore, the protocol defines the exchange of AI assets between actors.
This protocol specification is necessary as otherwise, every actor could use the smart contract differently, making it hard to use the information stored on the blockchain.
Finally, a client will be responsible for making these interactions accessible to a human actor.
All interactions performed by such a client will use the Ethereum account provided by the actor.

\subsection{Smart Contract}
\label{sub:smartcontract} 
The smart contract exposes the functionality that Clients will interact with.
For our use, the smart contract will only store information in immutable \textit{Logs} by executing events.
These logs are searchable and remain retrievable forever.
All information stored in logs this way is publicly accessible.
Additionally, blockchain technology does not allow to store large amounts of data directly on the blockchain.
Confidential data should, therefore, be encrypted and stored on an external storage solution as shown in Figure \ref{fig:systemOverview}.
Our protocol will introduce a way for exchanging such encrypted data that is provided through such storage solutions.

Table \ref{tab:smartcontractfunctions} shows all functions exposed by the smart contract.

\begin{table}
  \centering
  \caption{ Functions exposed on the smart contract, their description and required parameters.
  \label{tab:smartcontractfunctions}}
      \begin{tabularx}{\columnwidth}{@{}lX@{}}
          \hline
          \textbf{Function} & \textbf{Description} \\ \hline
          \textit{addAsset} & Register a new AI asset. The caller of the function will automatically be assigned as maintainer.\\
          \textit{transfer} & Transfer ownership of an AI asset from one blockchain user to another. This operation can only be performed by the maintainer of the asset.\\
          \textit{addUrl} & Add a download URL to an asset. This operation can only be performed by the maintainer of the asset.\\
          \textit{requestAccess} & Request access to an AI asset. This operation can be performed by any blockchain user.\\
          \textit{grantAccess} & Grant access to an AI asset. This operation can only be performed by the maintainer of the asset.\\
          \textit{getMaintainer} & Retrieve maintainer of an AI asset. This operation can be performed by any blockchain user.\\
      \end{tabularx}%
\end{table} 

A new AI asset is registered using the \textit{addAsset} function providing the following information:
An \textit{asset identifier} which is generated by computing a hash of the data item that is provided with the asset;
The \textit{URL} through which the data item of the asset can be retrieved;
Additional \textit{meta-information} that describes the contents of the asset;
This information needs to be provided as a JSON object.
The set of \textit{parents} that are in a relationship with this asset.
When the smart contract is executed, the \textit{maintainer} of the asset is automatically set to the user that called the function.

The execution of this function will trigger several events, which in turn will store the information as logs on the blockchain.
Listing \ref{lst:provenanceevents} shows the events related to the provenance tracking of AI assets.
When an AI asset is registered, the following events are emitted:
A \textit{Register} event that writes a log with the metadata of the asset, as well as a \textit{URL} event that stores the URL.
Lastly, all parents are written using \textit{ParentOf} events and the inverse \textit{ChildOf} events to ensure that the provenance graph remains in order and without cycles.

The \textit{FormerMaintainer} event is only used when the ownership of an AI asset changes.
Using these events allows us to store the complete provenance model to the blockchain, and therefore solve the privacy, auditability, and fair remuneration challenges.

\begin{lstlisting}[language=Solidity,numbers=none,caption={Provenance related events of the smart contract},captionpos=b,label={lst:provenanceevents}]
event Register(asset_id, metadata);
event URL(asset_id, url);
event FormerMaintainer(asset_id, previous_maintainer);
event ParentOf(asset_id, parent_id);
event ChildOf(asset_id, child_id);
\end{lstlisting}

To address the trust and confidentiality challenges, the smart contract needs to provide the functionality to exchange AI assets between actors in a confidential manner.
This works under the assumption that the data item, that can be downloaded from an AI asset URL, is encrypted.
The smart contract provides the two functions \textit{requestAccess} and \textit{grantAccess} that facilitate the exchange of the cryptographic information needed to decrypt the asset.
Listing \ref{lst:accessctrlevents} shows the additional events that store this information on the blockchain in the form of logs.
In the following section, we will introduce how the protocol specification facilitates this exchange.

\begin{lstlisting}[language=Solidity,numbers=none,caption={Event logs on the smart contract for the exchange of AI assets},captionpos=b,label={lst:accessctrlevents}]
event RequestAccess(asset_id, accessor, encryption_algorithm, public_key);
event GrantAccess(asset_id, accessor, encrypted_AEK);
\end{lstlisting}

\subsection{Protocol Specification}
\label{sub:protocol}
The smart contract specification from above defines a set of functionality that can be used by any application.
To ensure that all applications use the smart contract in the same fashion, we design an interaction protocol.
This will allow all systems implementing this protocol to rebuild the complete provenance of all AI assets stored on the blockchain.
Interactions with the smart contract are enabled through the JSON-RPC interface provided by the Ethereum network\footnote{\url{https://github.com/ethereum/wiki/wiki/JSON-RPC}}.

\begin{figure}[!t]
  \centering
  \subfloat[]{\includegraphics[width=.37\linewidth]{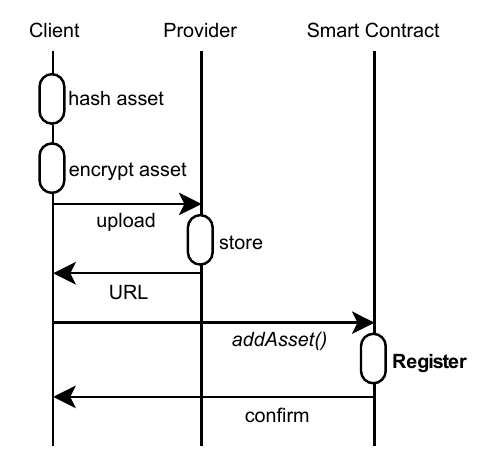}
  \label{fig:registration}}
  \hfil
  \subfloat[]{\includegraphics[width=.37\linewidth]{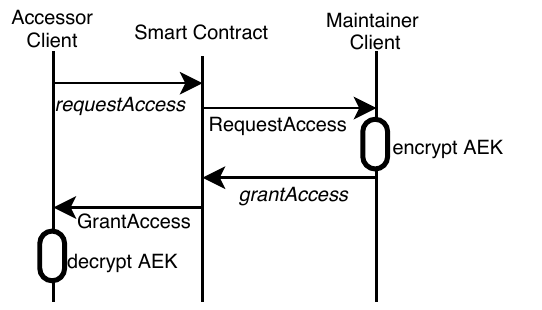}
  \label{fig:access}}
  \hfil
  \caption{
  Sequence diagrams of the registration (a) and the access retrieval (b) process.
  Function calls on the smart contract are in italics and emitted events in bold. 
  }
\end{figure}

\textbf{Registration}: Registering an AI asset is the key step to enable provenance for any asset.
Figure \ref{fig:registration} shows a sequence diagram of the protocol.
It will insert a new node into the provenance model and make it visible to others.
As already outlined, our system only supports assets that are made available through an external file storage provider.
Should the asset contain multiple files, it can be compressed into a ZIP file.
First, the Client computes a hash of the file to generate the asset identifier, and then encrypts it.
We call the encryption key used in this process the Asset Encryption Key (AEK) and it will be later used when access to an asset is granted.
Using the \textit{addAsset} function of the smart contract, the asset is then registered on the blockchain, and the account making this request will automatically become the maintainer of the asset.

\textbf{Accessing AI assets}: To solve the confidentiality challenge the data files representing an asset are encrypted as shown previously.
The smart contract and the protocol need to provide ways for exchanging the cryptographic material in a secure way that is registered on the blockchain.
Providing access to an AI asset needs an action of two actors.
First, the accessor creates an encryption key pair consisting of a private (PrK) and public key (PuK).
Then, they start the access process using the \textit{requestAccess} method of the smart contract, indicating the asset with its identifier.
A Client on the premises of the maintainer will react to the emitted \textit{RequestAccess} event.
If access should be granted the maintainer encrypts the AEK with the provided PuK, and invokes the \textit{grantAccess} method on the smart contract.
The Client of the accessor listens for the emitted \textit{GrantAccess} event containing the encrypted AEK that they will be able to decrypt using their PrK.
This then will allow them to download the asset from the external storage and decrypt it using the AEK.
The sequence diagram of this protocol is provided in Figure \ref{fig:access}.
Our system does not prevent misuse of the keys by the involved parties, e.g. the receiver simply providing the AEK to third-parties.
But the maintainer at least knows exactly whom they provided access to, limiting the number of potential bad actors.
It would also allow for the creation of systems that would automatically take the actual assets offline.

\textbf{Retrieve accessors}: 
We solve the outlined privacy challenge by making it possible for a maintainer to retrieve a list of actors who possibly accessed their AI assets.
As the smart contract emits the \textit{GrantAccess} event for all granted access requests, this can be solved easily. 
A client can retrieve and filter all past events using the Ethereum JSON-RPC API.
Using this information, a client can then visualize for each AI asset, which Ethereum accounts currently have access.
In the case of Data Erasure requests through the GDPR, the maintainer can then notify all users of such a data item to remove it as well.

\textbf{Retrieve Usages \& Build Provenance:} Finally, we need to provide a solution for the traceability and fair remuneration challenges.
We represent the potential usage of an AI asset if it is linked in a \textit{ParentOf} relationship to any other AI asset.
This also means that we can solve both challenges by building the provenance graph. 
As our provenance model is represented by an acyclic graph, we are able to compute it using existing graph traveling algorithms.
A Client can build the provenance graph by retrieving the \textit{ParentOf} events recursively, starting with the AI asset they are interested in.

This shows, that the protocol, combined with our generalized provenance model, addresses all challenges in Section \ref{sec:intro}.
\section{Validation} 
\label{sec:validation}
We validate our implementation by modeling the provenance of the \textit{Surgical Workflow Recognition for Collaborative Operation Theatre}~\cite{stauder2017} use case. 
The use case has been documented within the Bonseyes~\cite{llewellyn2017} project and reflects existing real-world use cases.
In this section, we first introduce the use case, its actors, actions, and involved AI assets.
We then translate this use case into the provenance model introduced in Section \ref{sec:provenancemodel}.
For simplicity, we abstract the use case and focus on the provenance affecting actions and the AI assets involved. 
We show which actors interact with the smart contract and present the resulting provenance model. 
Furthermore, we look into the costs of registering assets on the blockchain. 

\subsection{Medical Use Case}
\label{ss:tumusecase}
\begin{figure}[!t]
    \centering
    \subfloat[]{\includegraphics[width=.75\linewidth]{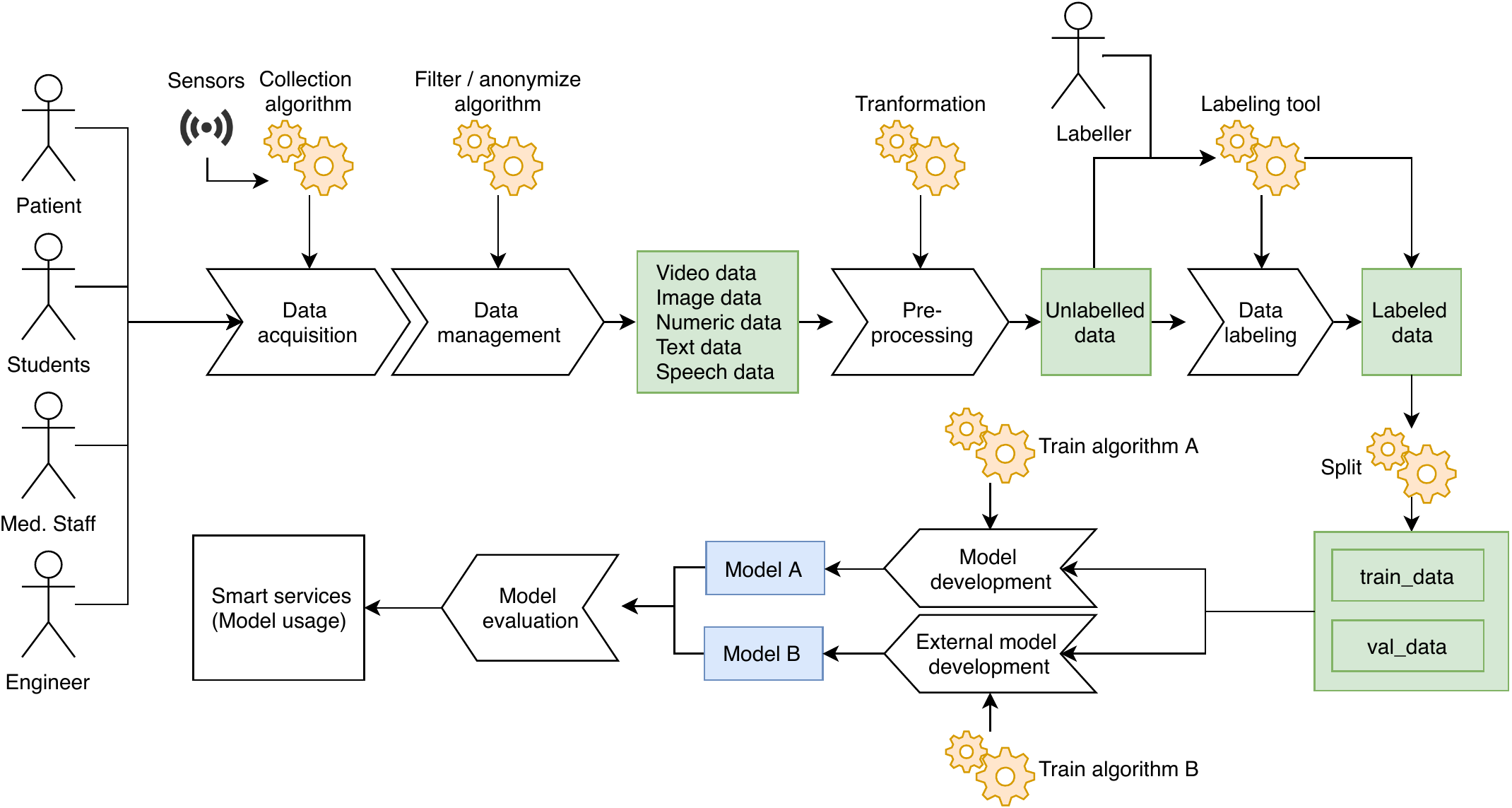}
    \label{fig:tumusecase}}
    \hfil
    \subfloat[]{\includegraphics[width=.24\linewidth]{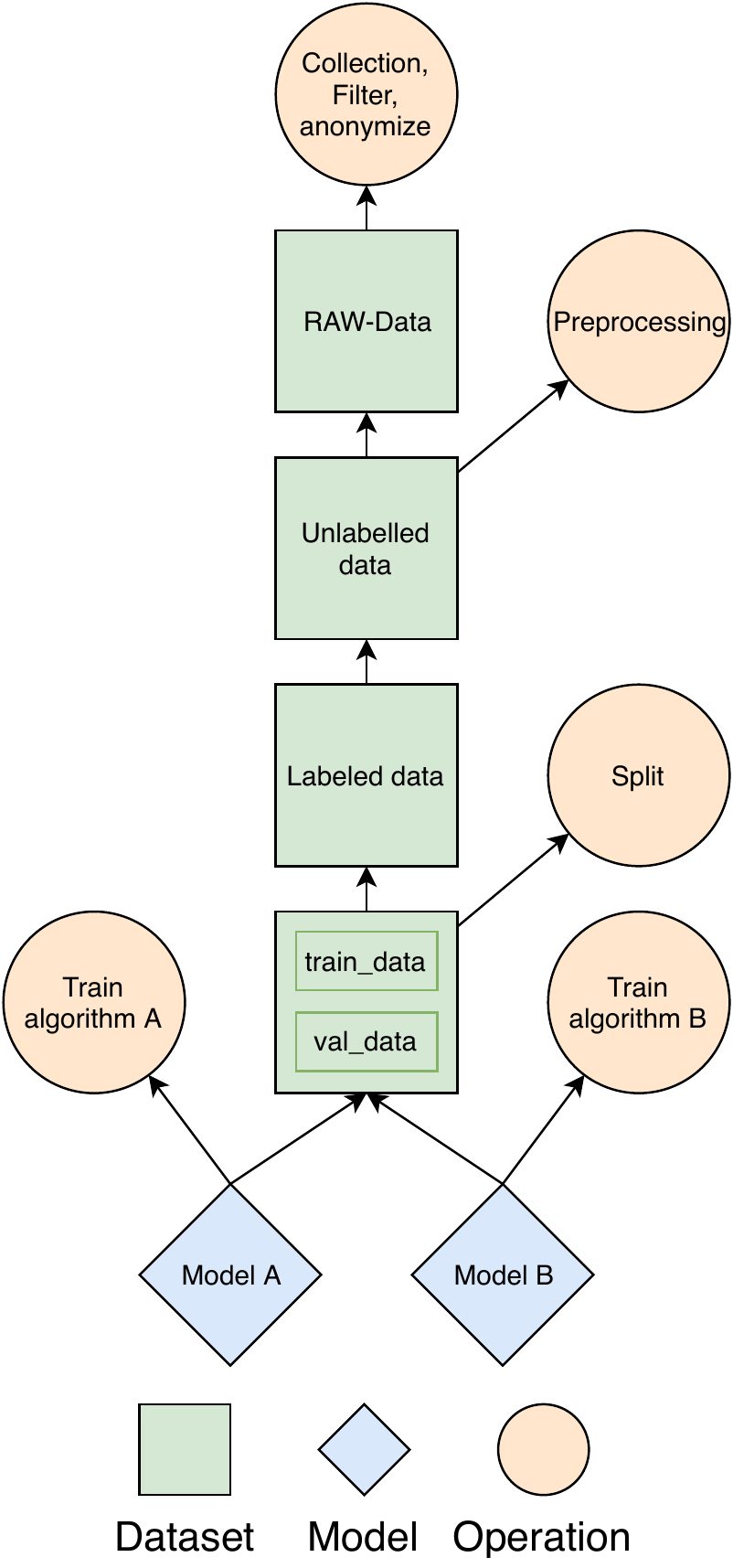}%
    \label{fig:tumusecaseprovenance}}
    \caption{Simplified Use Case: Surgical Workflow Recognition for collaborative Operation Theatre from Technical University Munich}
    \label{fig:tum}
\end{figure}

In Figure \ref{fig:tumusecase} we show the complete use case.
It covers all steps of a machine learning workflow from data collection to model deployment.
The goal of this use case is to develop AI models that support surgery inside the operation theatre.
Data is acquired from different sources in an operation theatre (e.g. cameras, and sensors).
The data collected during this phase is processed and managed by the data management action.
Different filter and anonymization algorithms support the data management action, all represented in a single action.
The result of this action is heterogeneous but anonymized RAW data.
Before labeling the data needs to be pre-processed.
This is performed by an algorithm that transforms and possibly aggregates the RAW data into a new dataset containing unlabeled data. 
This data is then fed into a labeling tool where expert labelers annotate the data.

As a result, we end up with a dataset viable for model development. 
Models are generated within the university hospital (e.g. by students) and externally by partners.
In order to be able to compare the generated models, the data is split into three parts (training, validation, and testing).
Only the training and validation datasets are made available to the development teams such that the testing dataset can be used for final evaluation.
For the external partners, the data leaves the network of the hospital for the first time.
Each development party uses its training algorithms that use the training data to generate an ML model.
These algorithms, as well as the resulting models, need to be registered using the smart contract.
Once the external parties provide their models back to the hospital, they again cross network boundaries.
For an audit of the training method, also the training algorithm might be exchanged.
The developed models will go through the model evaluation activity before they might get integrated into smart services.

\subsection{Smart Contract Interactions and Costs for TUM Use Case}
Along the outlined value chain, various AI assets are created and need to be registered on the smart contract.
As data acquisition and data management are performed in sequence by the same actor, we abstract these two actions into one.
In Figure \ref{fig:tumusecaseprovenance}, we show all AI assets identified in the value chain and the corresponding provenance graph.
Each of these assets is registered according to our protocol definition.
As in the real world, we use separate Ethereum accounts for every involved actor.

\begin{table}[ht]
  \caption{Gas cost in USD-Cents of registering and exchanging AI assets for each action of the TUM use case (Ether price: 210 USD, Gas price: 1E-9 Ether)}
  \label{tbl:tumgasconsuption}
  \centering
  \resizebox{\linewidth}{!}{%
  \begin{tabular}{@{}lrr@{}}
    \toprule
    Action & \multicolumn{1}{l}{Gas} & \multicolumn{1}{l}{Cost(\textcent)} \\ \midrule
    TUM registers data management algorithm & 74'669 & 15,7 \\
    TUM registers RAW data & 77'868 & 16,4 \\
    TUM registers unlabeled data and preprocessing algorithm. & 150'769 & 31,7 \\
    TUM registers now labeled data & 80'321 & 16,9 \\
    TUM registers split algorithm and train/val archive & 156'525 & 32,9 \\
    ExternalDataScientist requests access for archive & 72'573 & 15,2 \\
    TUM encrypts AEK for archive to grant access & 69'056 & 14,5 \\
    TUM registers own model and algorithm & 149'296 & 31,4 \\
    ExternalDataScientist registers their model and algorithm & 149'552 & 31,4 \\
    TUM requests access for Model B & 72'573 & 15,2 \\ \bottomrule
    \end{tabular}%
  }
  \end{table}

In Table \ref{tbl:tumgasconsuption}, we list the costs for each action that occurs inside the presented use case.
These actions correspond with the smart contract interactions presented in Table \ref{tab:smartcontractfunctions} using the protocol introduced in Section \ref{sub:protocol}.
Ether is the currency on the Ethereum blockchain and gas is the execution fee for an operation.
Each interaction with the smart contract costs around 0.15 USD.
The variations in costs are explained by the different amount of metadata that is provided for each AI asset.
Therefore, the complete provenance model of the outlined use case can be stored on the blockchain for less than 3 USD.

\subsection{Smart Contract Limitations}
As shown above, each execution of a smart contract on the Ethereum Network costs gas.
In order to prevent over-complex or non-terminating computations, the network limits the amount of gas a single transaction (or one function call) can use.
With the current gas limit, an AI asset can have a maximum of around 1200 parents.
In case of tracking complete datasets this might not be an issue, but if one aims at tracking single data points, e.g. for GDPR, this can quickly become an issue.
This would force data collectors to create multiple intermediate datasets, where data is incrementally aggregated.


A second limitation is that the smart contract stores metadata directly on the blockchain, thus making it publicly available.
All metadata of registered AI assets are therefore not confidential and accessible by anyone.
It remains an open discussion if there is a need for a distinction between required metadata that needs to be publicly available (e.g. for independent audits) and metadata information that can be made private.
\section{Discussion}
\label{sec:discussion}
Multiple experts from different companies collaborate in interacting AI value chains by exchanging AI assets.
Tracing the provenance of these AI assets from start to finish and especially across different AI value chains bears many challenges including trust, privacy, confidentiality, traceability and fair remuneration as defined in Section \ref{sec:intro}.

We introduce a graph-based provenance model generalizing the federated learning provenance model from Sarpatwar et al.~\cite{sarpatwar2019a} to support interacting value chains by solving the traceability, fair remuneration, and privacy challenges. 
Furthermore, we provide a smart contract for the Ethereum blockchain implementing the provenance model and removing the need for a middleman, thereby solving the trust challenge.
Additionally, the smart contract offers the ability to exchange assets in a confidential manner specified in a protocol fulfilling the confidentiality challenges.

Combining the provenance model and our protocol definition in a smart contract allowed us to track interacting value chains from start to finish.
We validated our smart contract with an industry use case from the Technical University of Munich, covering all phases of a typical machine learning workflow except the model monitoring phase.
Our results show that the smart contract is able to sufficiently trace the produced and exchanged AI assets at a low cost, as shown in Table \ref{tbl:tumgasconsuption}.

Comparing our concept to traditional centralized provenance tracing systems, we found that trust is a significant issue when collaborating.
Removing the centralized party with a decentralized blockchain increases trust among the participants and encourages collaboration.
Other blockchain-based provenance work, such as ProvChain~\cite{liang2017} and SmartProvenance~\cite{ramachandran2018}, exceed at tracking single file changes with sophisticated privacy features but cannot trace AI assets along all phases of value chains and are therefore not addressing the abovementioned challenges. 
The work of Sarpatwar et al.~\cite{sarpatwar2019a} focuses on enabling trusted AI for interacting value chains performing federated learning.
However, as the exchange of datasets is by design not supported, it is not possible to track interacting value chains outside of the federated learning context.
This gap is where our contribution steps in providing a solution for all challenges mentioned in Section \ref{sec:intro}.

As every transaction on a blockchain has its cost, our solution has its limitations.
We found that, with the current gas limit of ca. $7 \times 10^6$ on Ethereum, we can insert assets referencing up to $1200$ other AI assets as parents.
We consider this sufficient for most use cases.
Our smart contract was only tested on a local network discarding the waiting time that every transaction generates during block mining.
Furthermore, non-deterministic operations such as many AI training algorithms do not allow to reproduce the output asset completely and thus hinder auditability.
Finally, every client using our smart contract must implement our protocol to support the confidential exchange of assets as the smart contract is not able to enforce confidentiality without it.

For future work, it would be beneficial to test our smart contract with a sophisticated web client on the Ropsten or Ethereum network to include usability related aspects, such as the above mentioned waiting time when registering or exchanging assets.
Future extensions of the provenance model's type definitions, e.g. data streams, might allow increased provenance tracking coverage.
Furthermore, a comparison of our solution to a smart contract on a permissioned blockchain such as Hyperledger, which does not need any exchange management in the smart contract would help to decide if the blockchain or the smart contract should be responsible for the management of the AI asset exchange. 
Finally, zero-knowledge proofs may provide more secure auditability capabilities because access to assets would not be required.

\section*{Acknowledgements}
The project leading to this application has received funding from the European Unions Horizon 2020 research and innovation programme under grant agreement No 732204 (Bonseyes). 
This work is supported by the Swiss State Secretariat for Education, Research and Innovation (SERI) under contract numbers 16.0159. 
The opinions expressed and arguments employed herein do not necessarily reflect the official views of these funding bodies.
%
%
%
\bibliographystyle{splncs04}
\bibliography{bibliography}
\end{document}